# Skeletal editing by tip-induced chemistry

Shantanu Mishra,[1,2,*] Valentina Malave,[3,#] Rasmus Svensson,[1,4,#] Henrik Grönbeck,[1,4] Florian Albrecht,[2] Diego Peña,[3,5,*] Leo Gross[2,*]

[1]Department of Physics, Chalmers University of Technology, 412 96 Göteborg, Sweden

[2]IBM Research Europe – Zurich, 8803 Rüschlikon, Switzerland

[3]Center for Research in Biological Chemistry and Molecular Materials, and Department of Organic Chemistry, University of Santiago de Compostela, 15782 Santiago de Compostela, Spain

[4]Competence Centre for Catalysis, Chalmers University of Technology, 412 96 Göteborg, Sweden

[5]Oportunius, Galician Innovation Agency, 15702 Santiago de Compostela, Spain.

**ABSTRACT:** Skeletal editing of cyclic molecules has garnered considerable attention in the context of drug discovery and green chemistry, with notable examples in solution-phase synthesis. Here, we extend the scope of skeletal editing to the single-molecule scale. We demonstrate tip-induced oxygen deletion and ring contraction of an oxygen-containing seven-membered ring on bilayer NaCl films to generate molecules containing the perylene skeleton. The products were identified and characterized by atomic force and scanning tunneling microscopies, which provided access to bond-resolved molecular structures and orbital densities. Insights into the reaction mechanisms were obtained by density functional theory calculations. Our work expands the toolbox of tip-induced chemistry for single-molecule synthesis.

**Scheme 1. (a) Illustration of skeletal and peripheral editing. The solid arrows denote the skeletal editing reactions observed in this work. (b) Scheme showing the generation of the two majority products by tip-induced skeletal editing of DNO (left: atom deletion, right: ring contraction). Abundances of the two products are given relative to the total reaction obtained on the surface.**

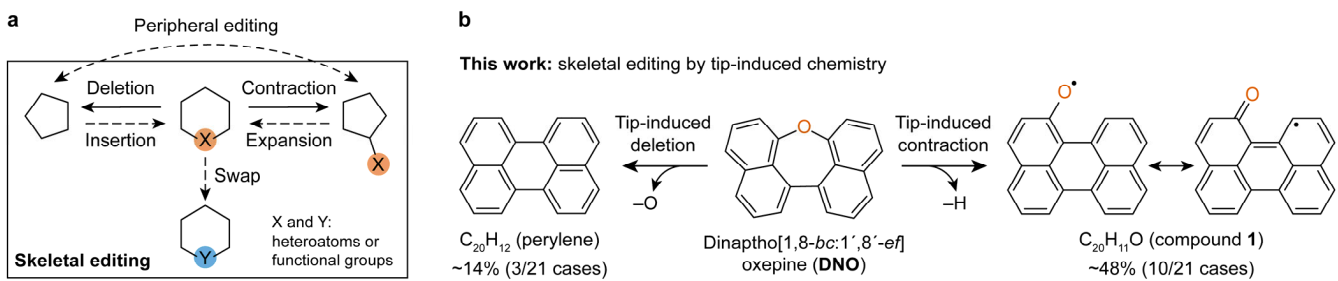

Skeletal editing[1] refers to the modification of the cyclic skeleton of an organic molecule by inserting, deleting or swapping individual atoms at precise locations at late stages of a synthetic sequence. The process is schematically depicted in Scheme 1, where it is contrasted with the more conventional peripheral editing (that is, functionalization), which does not modify the molecular skeleton but imparts changes at the molecular periphery. Skeletal editing is poised to be of great importance in medicinal chemistry,[2,3] where precise edits to the molecular skeleton without design of new synthetic routes from scratch could dramatically speed up drug discovery. Skeletal editing was also employed to modify polymer backbones,[4] which may open avenues for upcycling of plastics. The current repertoire of skeletal editing in solution-phase chemistry includes: (a) deletion[5–7] and insertion[8–10] of C, N ([14]N/[15]N) and O atoms; (b) swapping[11–13] of [12]C with [13]C, N and O; and (c) ring contraction[14] and expansion.[15] In this context, tip-induced chemistry,[16,17] wherein voltage pulses applied by a scanning probe tip are used to induce chemical reactions of molecular species adsorbed on surfaces, holds distinct advantages. The structural and electronic characterization of molecular species can be performed with atomic resolution by scanning tunneling (STM) and atomic force (AFM) microscopies. Furthermore, the solvent-free, cryogenic and ultra-high vacuum condition facilitates generation (sometimes with high selectivity[18]), stabilization and characterization of reactive intermediates and products, which can yield mechanistic insights into chemical reactions. In recent years, tip-induced chemistry has been successfully employed to generate molecules by peripheral editing, such as through dissociation of C–Cl,[18,19] C–Br,[20,21] C–I,[22] C–C,[23,24] C–H,[25,26] C–O,[27] N–H[28] and N–N[29] bonds, which may be followed by attachment of foreign molecules;[30] and by skeletal rearrangement.[31,32] Here, we demonstrate tip-induced skeletal editing, specifically, atom deletion and ring contraction (Scheme 1). From dinaphtho[1,8-*bc*:1′,8′-*ef*]oxepine (**DNO**), we generated different molecules with the perylene skeleton by oxygen deletion, and ring contraction and oxygen migration.

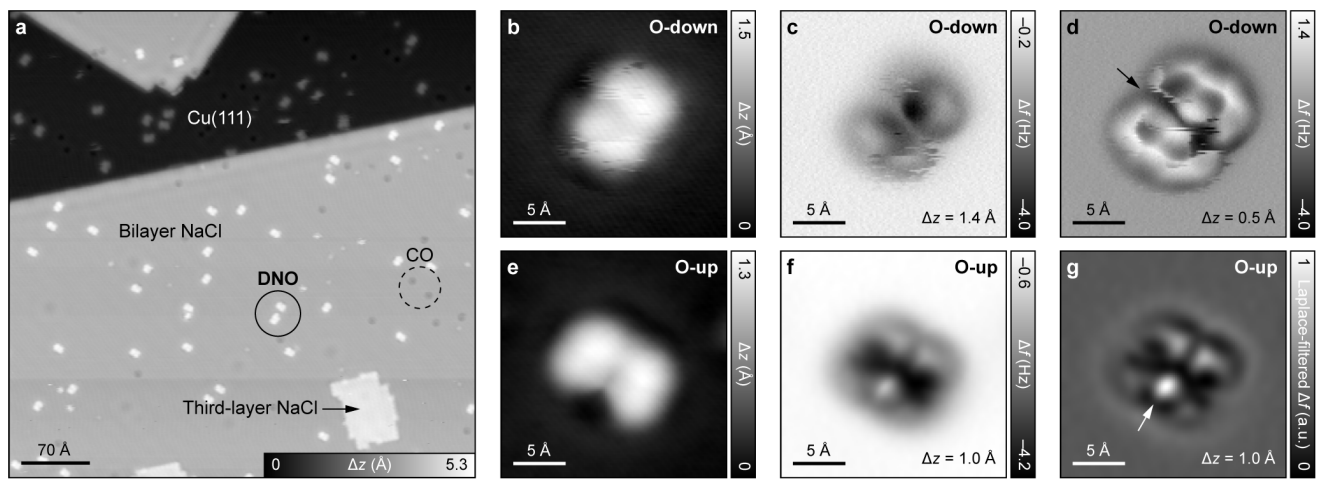

**Figure 1.** Characterization of **DNO** on NaCl/Cu(111). (a) STM overview image of the sample. (b–d) STM (b) and AFM (c,d) images of an O-down **DNO** molecule. (e–g) STM (e), AFM (f) and the corresponding Laplace-filtered AFM (g) images of an O-up **DNO** molecule. The arrows in (d, g) indicate the location of the oxygen atom. Scanning parameters for STM images: $V = 0.2$ V, $I = 0.5$ pA. Open feedback parameters for AFM images: $V = 0.2$ V, $I = 0.5$ pA on NaCl. $\Delta z$ denotes the tip-height offset, with positive (negative) values indicating tip retraction (approach) from the set-point value.

**DNO** was synthesized by solution-phase chemistry (Schemes S1–S5 and Figs. S1-S7) and deposited on a single-crystal Cu(111) surface partially covered by bilayer NaCl films (Fig. S8). The crystal was contained within a combined STM/AFM apparatus operating under ultra-high vacuum and at a temperature of 5 K. Figure 1a shows an STM image of the surface, revealing isolated **DNO** molecules, along with co-adsorbed carbon monoxide (CO) molecules and a minority of third-layer NaCl islands. Nearly all **DNO** molecules on NaCl adopt an adsorption conformation wherein the oxygen atom points toward the surface (denoted O-down). Figures 1b–d show STM and AFM images of an O-down **DNO** molecule. In AFM imaging at large tip heights (Fig. 1c), only two benzenoid rings of the molecule are visible, and in imaging at small tip heights all four benzenoid rings become visible (Fig. 1d). Related to the adsorption geometry, the oxygen atom is not visible. The conspicuous streaks in the STM and AFM images likely result from conformational switching of the molecule under the influence of the tip, related to the relative tilting of the benzenoid rings. Rarely, we found **DNO** molecules on NaCl wherein the oxygen atom points away from the surface (O-up, Figs. 1e–g). In this case, the oxygen atom, being the atomic species closest to the tip, appears bright in AFM imaging (that is, with an increased frequency shift, $\Delta f$). Density functional theory (DFT) calculations of **DNO** on NaCl/Cu(111) predict the O-down conformation to be ~0.1 eV more stable than the O-up, in line with the experimentally observed predominance of O-down species. Figure S9 shows further measurements on **DNO**.

Voltage pulses ranging between 4.9–5.1 V were applied to individual **DNO** molecules by the tip of the STM/AFM system to trigger intramolecular chemical reactions (see Supporting Information). The reaction yield with such pulses was ~21%; that is, for 99 voltage pulses applied to 28 **DNO** molecules, 21 molecules underwent a reaction on the surface, resulting in 6 unique species. In unsuccessful cases, **DNO** molecules were either displaced on NaCl and remained intact, or were not located, presumably being picked up by the tip or displaced from NaCl onto Cu(111). Out of the 21 reacted molecules, the majority (16/21) corresponded to molecules with the perylene skeleton (constituting 4 species); namely, perylene (3/21), 1-perylenoxy radical (compound **1**; 10/21), perylenyl radical (compound **2**; 2/21) and didehydroperylene (compound **3**; 1/21). In cases where the oxygen atom was removed from **DNO** (as in perylene, **2** and **3**), the oxygen atom not found on the surface, likely because of its desorption to the gas phase or diffusion on NaCl followed by adsorption on Cu(111). Here, we focus on the characterization of perylene and **1**, the major products of atom deletion and ring contraction, respectively. The characterizations of **2** and **3** are reported in Fig. S10, and data on the remaining two species (5/21 cases) are shown in Fig. S11.

Figure 2a shows an AFM image of perylene generated by tip-induced removal of an oxygen atom from **DNO**. STM imaging at bias voltage $V = 1.8$ V (Fig. 2b) revealed the lowest unoccupied molecular orbital (LUMO) density of perylene, in agreement with calculations (Fig. S12). We next focus on the characterization of **1**, which maintains an oxygen atom attached to the perylene skeleton (Scheme 1). At the outset, we note that **1** is found in an anionic charge state on the surface (denoted **1**⁻). This is experimentally inferred from the scattering of the NaCl/Cu(111) interface state by the molecule (Fig. S13) and supported by Bader charge analysis of the molecule adsorbed on bilayer NaCl/Cu(111), yielding a transfer of 0.81 electrons from the surface to **1**. In contrast to charge-neutral **1** (denoted **1⁰**) being an open-shell species containing an unpaired π-electron, **1⁻** is closed-shell (Fig. S14). The optimized C–O bond length of **1⁻** on bilayer NaCl/Cu(111) is calculated to be 1.29 Å. To contextualize this value in terms of the single or double bond character of the C–O bond, we carried out gas-phase calculations (Table S1) of cyclohexanone (containing a C($sp^2$)–O double bond) and phenol (containing a C($sp^2$)–O single bond). From these calculations, we obtained C–O bond lengths of 1.23 (cyclohexanone) and 1.38 Å (phenol). The comparison of the C–O bond lengths thus indicates that **1⁻** is best described as a resonance hybrid of structures containing both C–O single and double bonds (Scheme 1). Figures 2c–e show the STM and AFM images of **1⁻**. In AFM imaging of **1⁻** (Figs. 2d, e), the salient feature is the notably dark appearance of the benzenoid ring bonded to an oxygen atom (highlighted by the arrow in Fig. 2d), along with the bright appearance of two peripheral benzenoid rings that lie diagonally opposite to each other. This observation agrees with the DFT calculations of **1⁻** on bilayer NaCl/Cu(111) (Fig. 2f), which shows a non-planar adsorption conformation of the molecule with the oxygen atom pointing toward the surface, and an upward tilt of the two peripheral benzenoid rings. Figures 2g, h show the STM and AFM images of **1⁻** adsorbed next to a **DNO** molecule that rendered **1⁻** stable enough to be imaged at elevated voltages. We could therefore measure the HOMO density of **1⁻** at −1.0 V (corresponding to the detachment of an electron from **1⁻**, Fig. 2i), which exhibited good agreement with the calculated HOMO of **1⁻** (Fig. 2j).

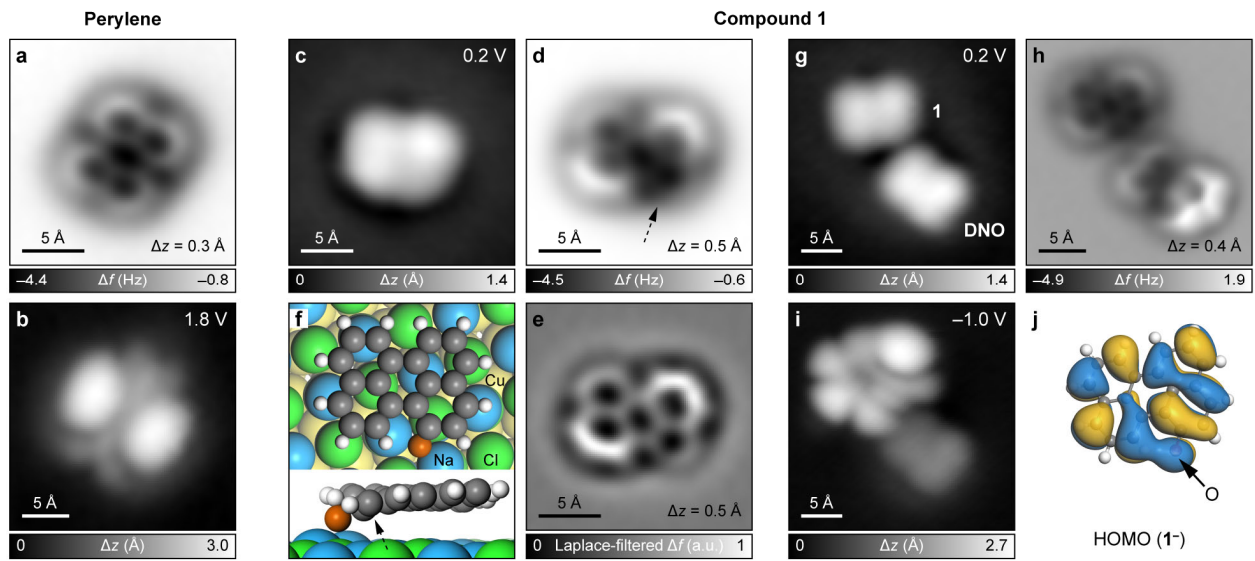

**Figure 2.** Structural and electronic characterization of perylene and compound **1**. (a) AFM image of perylene. (b) STM image of perylene showing its LUMO density. (c–e) STM (c), AFM (d) and corresponding Laplace-filtered AFM (e) images of **1⁻**. (f) Top and side views of the DFT calculated adsorption geometry of **1⁻** on bilayer NaCl/Cu(111). The oxygen atom adsorbs bridging two Na⁺ ions, which show an outward relaxation of ∼0.5 Å. The arrows in (d, f) indicate the six-membered ring bonded to the oxygen atom. C, H and O atoms are colored gray, white and orange, respectively. (g, h) STM (g) and AFM (h) images of **1⁻** adsorbed next to a **DNO** molecule. (i) Corresponding STM image showing the HOMO density of **1⁻**. (j) DFT calculated HOMO of **1⁻** (isosurface: $0.01a_0^{-3/2}$, $a_0$ denotes the Bohr radius). Scanning parameters for STM images: $I$ = 0.15 pA (b), $I$ = 0.5 pA (c), and $I$ = 0.3 pA (g, i). Open feedback parameters for AFM images: $V$ = 0.2 V, $I$ = 0.5 pA on NaCl.

Insights into the mechanisms of skeletal editing reactions were obtained by DFT calculations of the reaction landscape. Figure 3 summarizes the results and highlights the relevant reaction paths and energy barriers involved in the generation of perylene, **1**, and **2**. Because of the weak molecule-surface interactions on NaCl, we focus on the discussion of the reaction mechanisms in the gas phase (solid lines in Fig. 3). Starting from **DNO**, the first step involves the formation of a C-C bond resulting in a central six-membered ring, along with migration of the oxygen atom to a neighboring C–C bridge site (epoxyperylene intermediate **Int1**, activation energy $\Delta E^{\ddagger}$ = 2.70 eV). Three reaction paths are possible after the first step. First, the oxygen atom in **Int1** can be eliminated ($\Delta E^{\ddagger}$ = 3.85 eV), leading to the generation of perylene. Second, the oxygen atom in **Int1** can migrate to the next C–C bridge site ($\Delta E^{\ddagger}$ = 1.02 eV) leading to the epoxyperylene intermediate **Int2**, from where the oxygen atom can once again be eliminated ($\Delta E^{\ddagger}$ = 3.44 eV) to generate perylene. Third, a rearrangement reaction can occur, resulting in the formation of intermediate **Int3** (perylen-1-ol). The reaction is exothermic and associated with $\Delta E^{\ddagger}$ = 1.20 eV. From **Int3**, either the hydroxy group can be removed to generate **2**, which requires a large $\Delta E^{\ddagger}$ of 4.81 eV, or the hydrogen atom can be removed from the hydroxy group to generate **1** ($\Delta E^{\ddagger}$ = 3.28 eV). We also calculated the potential energy landscape on bilayer NaCl/Cu(111) (dashed lines in Fig. 3), where the relative energies of all the intermediates and products are similar to the gas phase, apart from **1**, which is stabilized on the surface (Fig. 2f). We note that intermediates **Int1–Int3** were not observed in the experiments. The gas-phase calculations in Fig. 3 were performed assuming a neutral charge state of the reactant, intermediates and products. We also performed gas-phase calculations assuming a global anionic state, which is a possibility given the large positive applied voltage pulses that could cause transient charging of the species. As shown in Fig. S15, the reaction mechanisms in the anionic state are similar, but there is a notable lowering of all activation barriers compared to the neutral case. From experiments, we cannot deduce whether the reactions occur in the neutral or anionic charge state. Compound **1** might be charged by electron transfer from the surface after it is formed. The importance of using the insulating NaCl surface to reduce molecule-surface interactions and facilitate skeletal editing is emphasized by the fact that we were unable to perform skeletal editing on Cu(111), where chemisorption of intermediates or competing reactions such as C–H bond activation were observed instead (Fig. S16).

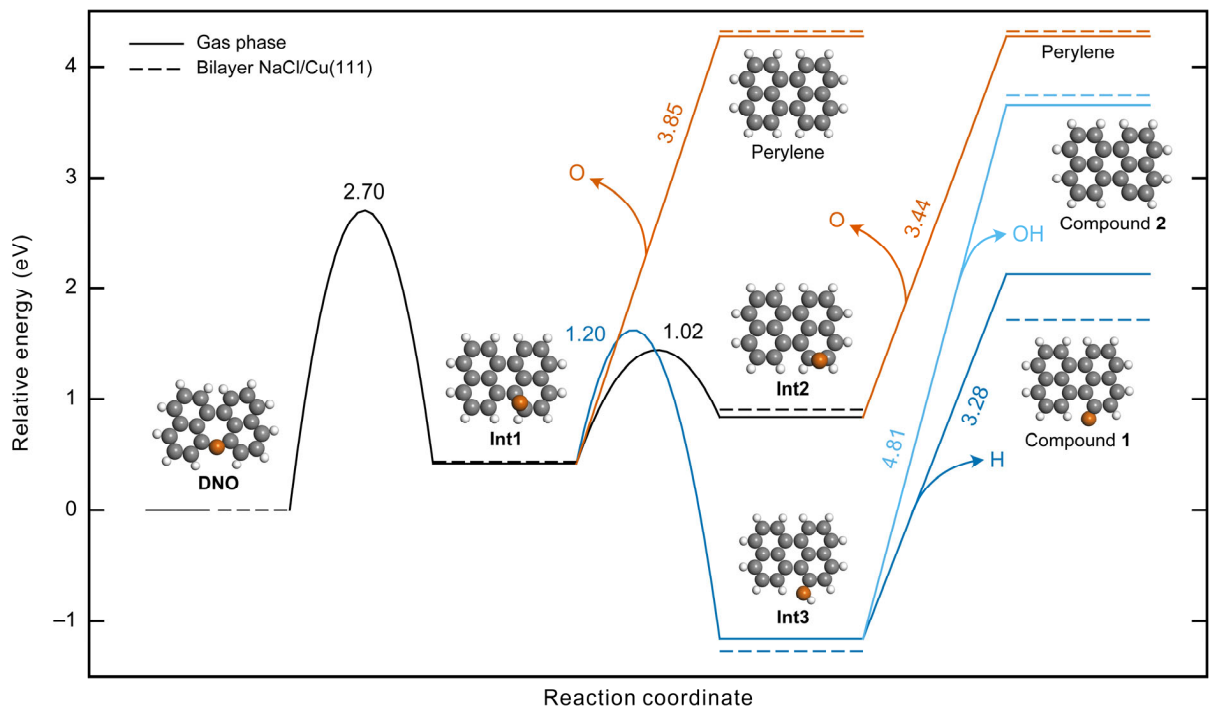

**Figure 3.** Ab-initio potential energy landscape in the gas phase (solid lines) and on bilayer NaCl/Cu(111) (dashed lines) for skeletal editing of **DNO**. The optimized gas-phase geometries of **DNO**, intermediates and products are also shown. The numbers adjacent to the curves denote absolute gas-phase activation energies in eV. Note that for the reactions **Int1** and **Int2** → perylene and **Int3** → compounds **1** and **2**, energy differences coincide with activation energies. For on-surface calculations, gas-phase barrier heights are used; and the dissociated O, H and OH species are assumed to desorb.

In conclusion, we demonstrate skeletal editing by tip-induced chemistry. Voltage pulses applied by a scanning probe tip to dinaptho[1,8-*bc*:1′,8′-*ef*]oxepine molecules adsorbed on NaCl resulted in two distinct skeletal editing reactions – atom deletion and ring contraction. The former predominantly led to the generation of perylene, whereas the latter led to the generation of 1-perylenoxy radical. Experimental atomic-scale characterization of the products was performed by STM and AFM, and a detailed mechanistic understanding of the intramolecular reactions leading to the observed products was obtained by DFT calculations. Future directions could involve exploring how heteroatoms of the same group (such as sulfur) will affect the reaction landscape and yields, and performing the challenging atom insertion and swap edits by tip-induced chemistry. Obtaining selectivity of different skeletal editing reactions by tip-induced chemistry would be another goal.

The addition of skeletal editing to the toolbox of tip-induced chemistry provides a powerful route for single-molecule synthesis. Moreover, tip-induced skeletal editing may be used for precise local modification of heteroatom-containing carbon nanostructures to imprint functionalities such as semiconducting heterojunctions,[33,34] magnetism[35] and topological electronic phases.[36,37]

## ASSOCIATED CONTENT

**Supporting Information**. Experimental and theoretical methods, solution synthesis of **DNO**, [1]H and [13]C NMR spectra, mass spectroscopy data, STM and AFM data, and additional calculations.

## AUTHOR INFORMATION

### Author Contributions

#These authors contributed equally.

### Corresponding Authors

*__Shantanu Mishra__ – shantanu.mishra@chalmers.se,
*__Diego Peña__ – diego.pena@usc.es,
*__Leo Gross__ – LGR@zurich.ibm.com.

### Funding Sources

The European Research Council Synergy grant MolDAM (grant number 951519), the Spanish Agencia Estatal de Investigación (grant number PID2022-1408450B-C62), Xunta de Galicia (Centro de Investigación do Sistema Universitario de Galicia, 2023–2027, grant number ED431G 2023/03), the European Regional Development Fund, and the Swedish Research Council (grant number 2024-05250). The calculations were performed at PDC and NSC via a NAISS grant.

Notes
The authors declare no competing financial interests.

ACKNOWLEDGMENTS

The authors thank Katja-Sophia Csizi and Lisanne Sellies for discussions.

# Supporting Information

## Skeletal editing by tip-induced chemistry

Shantanu Mishra,[1,2] Valentina Malave,[3] Rasmus Svensson,[1,4] Henrik Grönbeck,[1,4] Florian Albrecht,[2] Diego Peña,[3,5] Leo Gross[2]

[1]Department of Physics, Chalmers University of Technology, 412 96 Göteborg, Sweden

[2]IBM Research Europe – Zurich, 8803 Rüschlikon, Switzerland

[3]Center for Research in Biological Chemistry and Molecular Materials, and Department of Organic Chemistry, University of Santiago de Compostela, 15782 Santiago de Compostela, Spain

[4]Competence Centre for Catalysis, Chalmers University of Technology, 412 96 Göteborg, Sweden

[5]Oportunius, Galician Innovation Agency (GAIN), 15702 Santiago de Compostela, Spain

**Contents**

# 1. Methods

## 1.1. Solution synthesis and characterization

Starting materials were purchased reagent grade from TCI or Sigma-Aldrich and used without further purification. $CH_2Cl_2$ was dried using a MBraun SPS-800 Solvent Purification System. All reactions were carried out in flame-dried glassware under an inert atmosphere of purified Ar using Schlenk techniques. Deuterated solvents were purchased from Acros Organics. Thin-layer chromatography (TLC) was performed on Silica Gel 60 F-254 plates (Merck) and chromatograms were visualized with UV light (254 and 365 nm) and/or stained with Hanessian's stain. Column chromatography was performed on silica gel (40-60 μm). $^1H$ and $^{13}C$ NMR spectra were recorded at 300 MHz ($^1H$) with Varian Mercury 300 instrument, and at 500 MHz ($^1H$) and 125 MHz ($^{13}C$) MHz with Bruker 500 instrument. Mass spectra, using the atmospheric pressure chemical ionization (APCI) method, were recorded on a Bruker MicroTOF spectrometer.

Oxepin **DNO** was obtained following the route shown in Scheme S1, which is based on a previously reported procedure.[1]

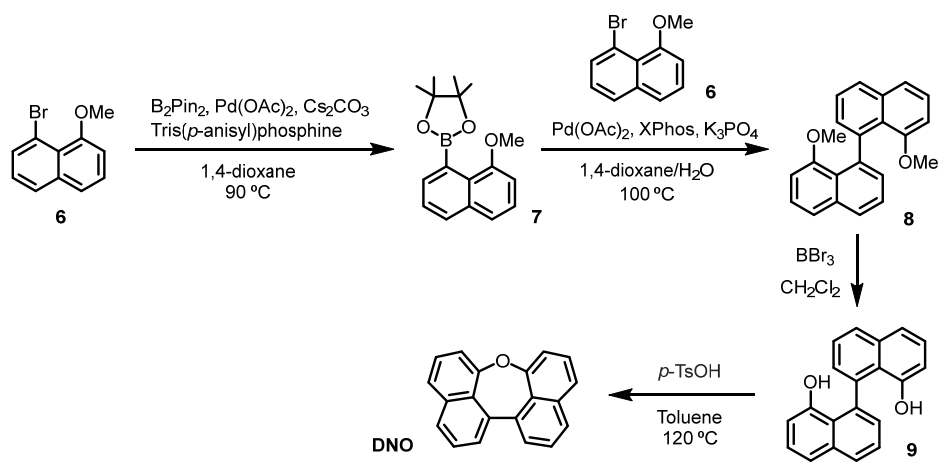

**Scheme S1**. Synthetic route toward the oxepin **DNO**.

## Synthesis of compound 7

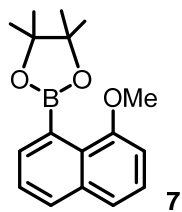

**Scheme S2.** Chemical structure of compound **7**.

To a flame-dried Schlenk flask, compound **6** (100 mg, 0.42 mmol), bis(pinacolato)diboron ($B_2Pin_2$) (160.7 mg, 0.63 mmol), $Cs_2CO_3$ (412.3 mg, 1.26 mmol), Pd(OAc)$_2$ (9.5 mg, 0.04 mmol) and tris(4-methoxyphenyl)phosine (29.7 mg, 0.08 mmol) were added. Then, 1,4-dioxane (6.5 mL) was added, and the reaction mixture was stirred at 90 ºC for 5 h. After reaction completion (monitored by TLC), the reaction mixture was cooled to room temperature and concentrated under reduced pressure to remove the solvent. The crude residue was purified by column chromatography (SiO₂; hexane:CH₂Cl₂ 3:2) to afford compound **7** (107 mg, 89%) as a white solid.

**¹H NMR** (500 MHz, CDCl₃, Fig. S1) δ: 7.79 (dd, *J* = 8.1 Hz, 1H), 7.52 (dd, *J* = 6.8 Hz, 1H), 7.47 – 7.40 (m, 2H), 7.38 – 7.34 (m, 1H), 6.84 (dd, *J* = 7.6 Hz, 1H), 4.01 (s, 3H), 1.45 (s, 12H) ppm. **¹³C NMR** (125 MHz, CDCl₃, Fig. S2) δ: 155.77 (C), 134.31 (C), 130.18 (CH), 128.74 (CH), 127.57 (C), 125.74 (CH), 121.17 (CH), 104.75 (CH), 83.72 (C), 55.70 (CH₃), 25.28 (CH₃) ppm. **MS (APCI)** *m/z* (%): 283 (M-1, 100) (MS denotes mass spectrum).

## Synthesis of compound 8

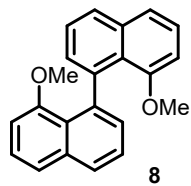

**Scheme S3.** Chemical structure of compound **8**.

To a round-bottom flask, compound **7** (107.0 mg, 0.38 mmol), compound **6** (50.0 mg, 0.21 mmol), K₃PO₄ (134.3 mg, 0.63 mmol), Pd(OAc)₂ (9.5 mg, 0.04 mmol) and XPhos (20.1 mg, 0.04 mmol) were added. The solids were dissolved in a mixture of 1,4-dioxane/H₂O (1:1, 4.0 mL), and the resulting solution was heated at 100 ºC for 5 h. After reaction completion (monitored by TLC), the reaction mixture was cooled to room temperature and diluted with CH₂Cl₂. The organic phase was washed with water, dried over Na₂SO₄, filtered, and concentrated under reduced pressure. The crude residue was purified by column chromatography (SiO₂; hexane:CH₂Cl₂ 1:1) to afford compound **8** (61.0 mg, 91%) as a white solid.

**¹H NMR** (500 MHz, CDCl₃, Fig. S3) δ: 7.78 (dd, *J* = 8.2 Hz, 2H), 7.50 (d, *J* = 8.1 Hz, 2H), 7.47 – 7.42 (m, 2H), 7.34 (t, *J* = 7.9 Hz, 2H), 7.21 (dd, *J* = 7.0 Hz, 2H), 6.66 (d, *J* = 7.5 Hz, 2H), 3.05 – 3.03 (m, 6H) ppm. **¹³C NMR** (125 MHz, CDCl₃, Fig. S4) δ: 157.42 (C), 142.19 (C), 134.86 (C), 126.84 (CH), 126.46 (CH), 125.65 (CH), 125.42 (CH), 125.16 (CH), 121.06 (CH), 105.92 (CH), 55.52 (CH₃) ppm. **MS (APCI)** *m/z* (%): 315 (M+1, 100).

## Synthesis of compound 9

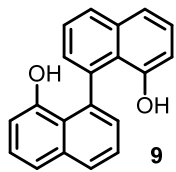

**Scheme S4.** Chemical structure of compound **9**.

To a solution of compound **8** (75.0 mg, 0.24 mmol) in CH₂Cl₂ (2.0 mL) at –78 ºC, BBr₃ (80.0 µL, 0.84 mmol) was added dropwise. After the addition, the reaction mixture was allowed to warm to room temperature and stirred for 2 h. The reaction was then quenched by careful addition of water, and the mixture was extracted with CH₂Cl₂. The combined organic fractions were washed with water, dried over Na₂SO₄, filtered and concentrated under reduced pressure. The crude product was purified by column chromatography (SiO₂; hexane:CH₂Cl₂ 1:1) to afford compound **9** (65.0 mg, 95%) as a white solid.

**¹H NMR** (500 MHz, CDCl₃, Fig. S5) δ: 7.98 (dd, *J* = 8.3, 1.3 Hz, 2H), 7.56 (dd, *J* = 8.2, 1.2 Hz, 2H), 7.51 (t, *J* = 7.7 Hz, 2H), 7.44 (t, *J* = 7.9 Hz, 2H), 7.38 (dd, *J* = 7.0, 1.3 Hz, 2H), 6.88 (dd, *J* = 7.6, 1.2 Hz, 2H),

5.39 (s, 2H) ppm. **$^{13}$C NMR** (125 MHz, CDCl$_3$, Fig. S6) δ: 153.36 (C), 135.89 (C), 134.75 (C), 130.21 (CH), 128.90 (CH), 127.89 (CH), 125.01 (CH), 122.11 (C), 121.41 (CH), 112.74 (CH) ppm. **MS (APCI)** *m/z* (%): 286 (M+1, 100).

**Synthesis of DNO**

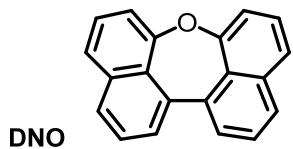

**DNO**

**Scheme S5.** Chemical structure of **DNO**.

Compound **9** (64.0 mg, 0.22 mmol) and *p*-toluenesulfonic acid (63.8 mg, 0.33 mmol) were dissolved in toluene (2.3 mL), and the reaction mixture was refluxed for 3 h. After cooling to room temperature, the solution was quenched with saturated aqueous solution of K$_2$CO$_3$ and the phases were separated. The organic layer was washed with water, dried over MgSO$_4$, filtered, and concentrated under reduced pressure. The crude solid was recrystallized from *n*-hexane to afford **DNO** (26.0 mg, 43%) as a yellow solid.

**$^1$H NMR** (300 MHz, CDCl$_3$, Fig. S7) δ: 8.19 (d, *J* = 7.6 Hz, 1H), 7.80 (d, *J* = 8.0 Hz, 1H), 7.66 (dd, *J* = 6.7, 2.7 Hz, 1H), 7.44 (t, *J* = 7.9 Hz, 2H), 7.38 (dd, *J* = 7.0, 1.3 Hz, 2H), 6.88 (dd, *J* = 7.6, 1.2 Hz, 2H), 7.58 – 7.42 (m, 3H) ppm. **MS (APCI)** *m/z* (%): 268 (M+1, 100).

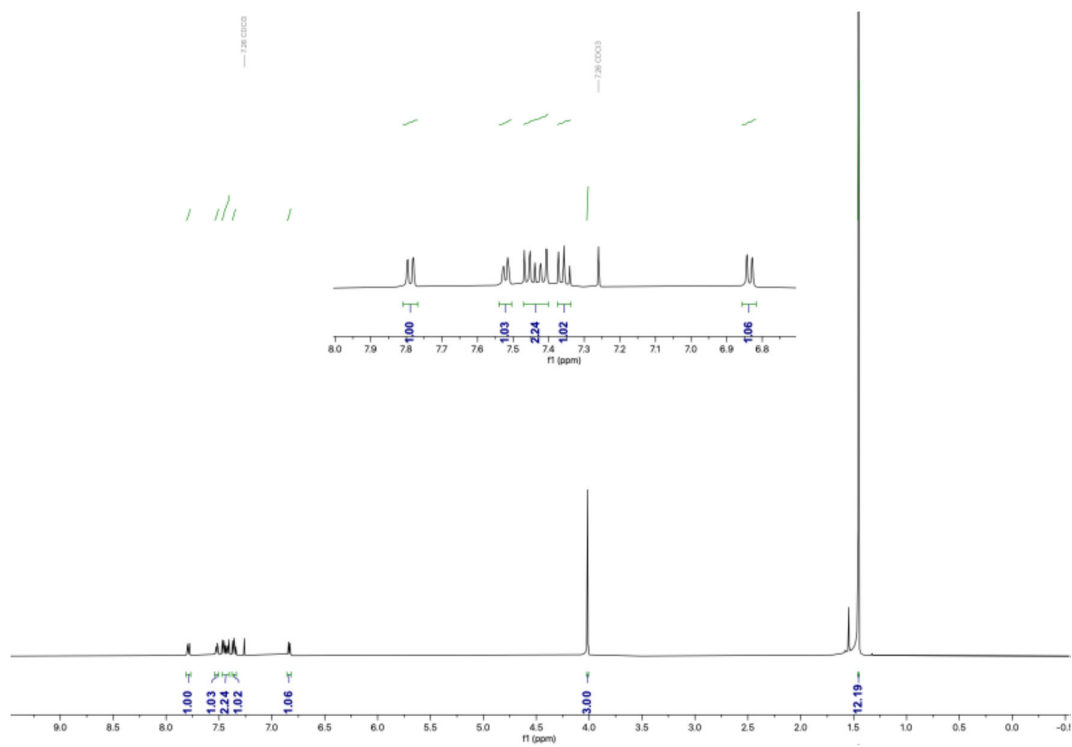

**Figure S1**. ¹H NMR (500 MHz, CDCl₃) spectrum of compound **7**.

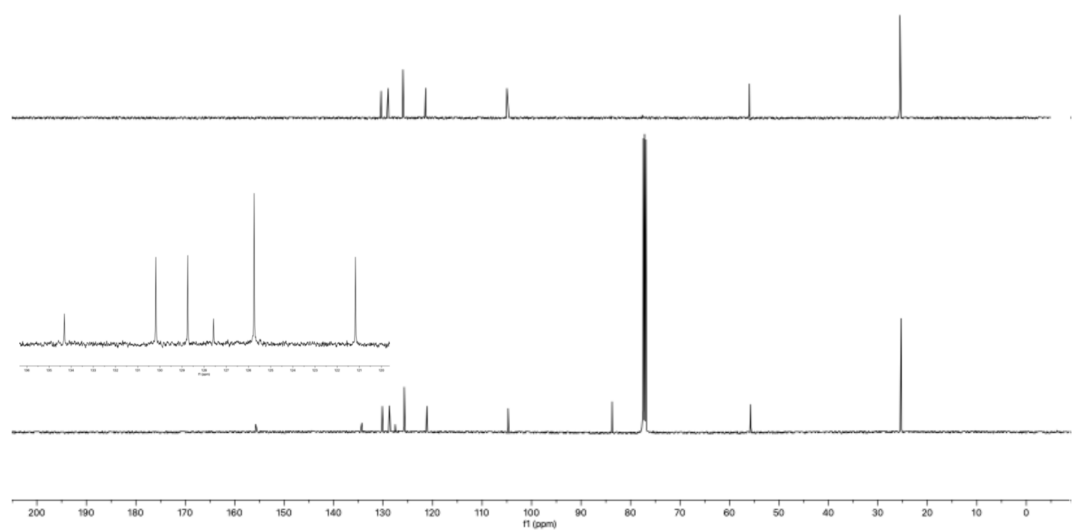

**Figure S2**. ¹³C NMR-DEPT (125 MHz, CDCl₃) spectrum of compound **7**.

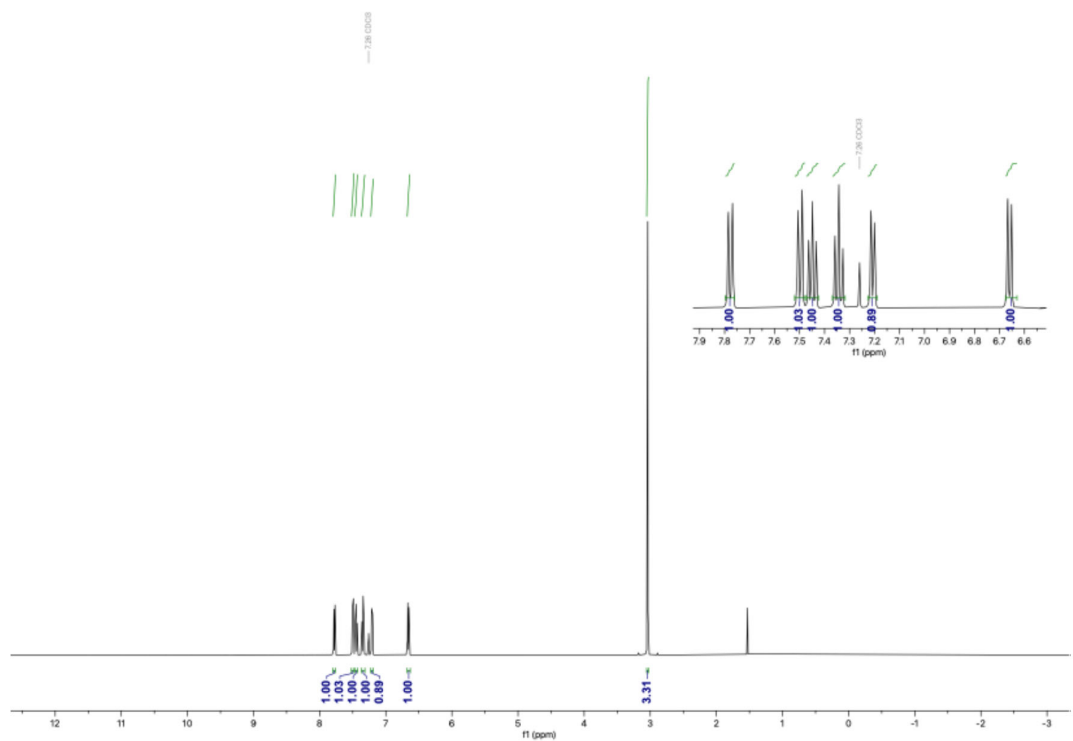

**Figure S3**. $^1$H NMR (500 MHz, CDCl$_3$) spectrum of compound **8**.

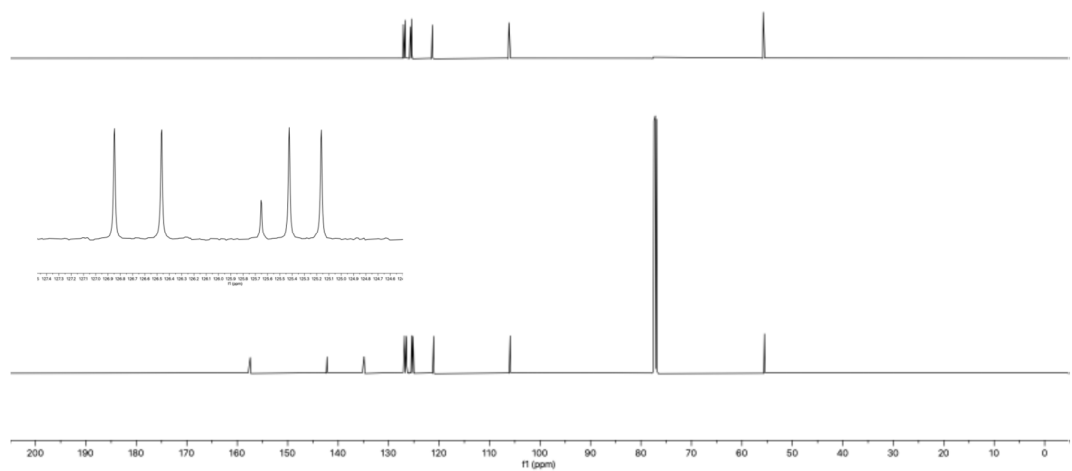

**Figure S4**. $^{13}$C NMR-DEPT (125 MHz, CDCl$_3$) spectrum of compound **8**.

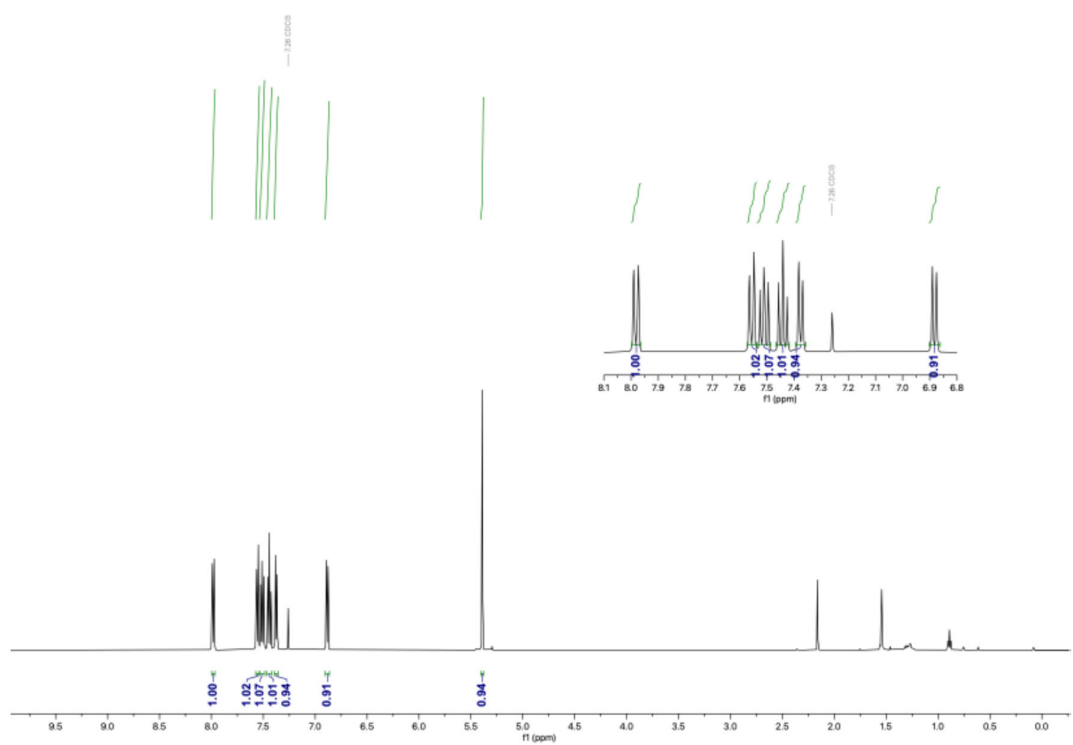

**Figure S5**. ¹H NMR (500 MHz, CDCl₃) spectrum of compound **9**.

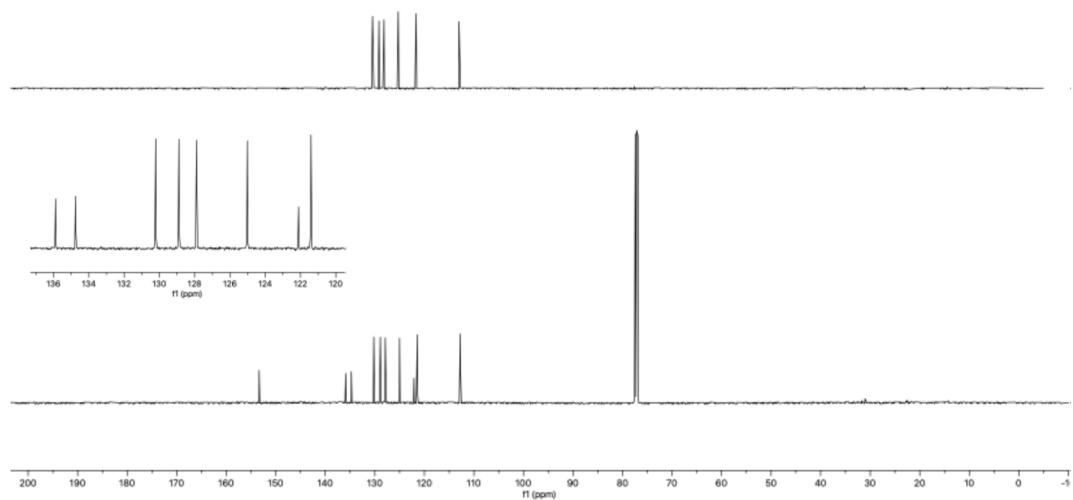

**Figure S6**. ¹³C NMR-DEPT (125 MHz, CDCl₃) spectrum of compound **9**.

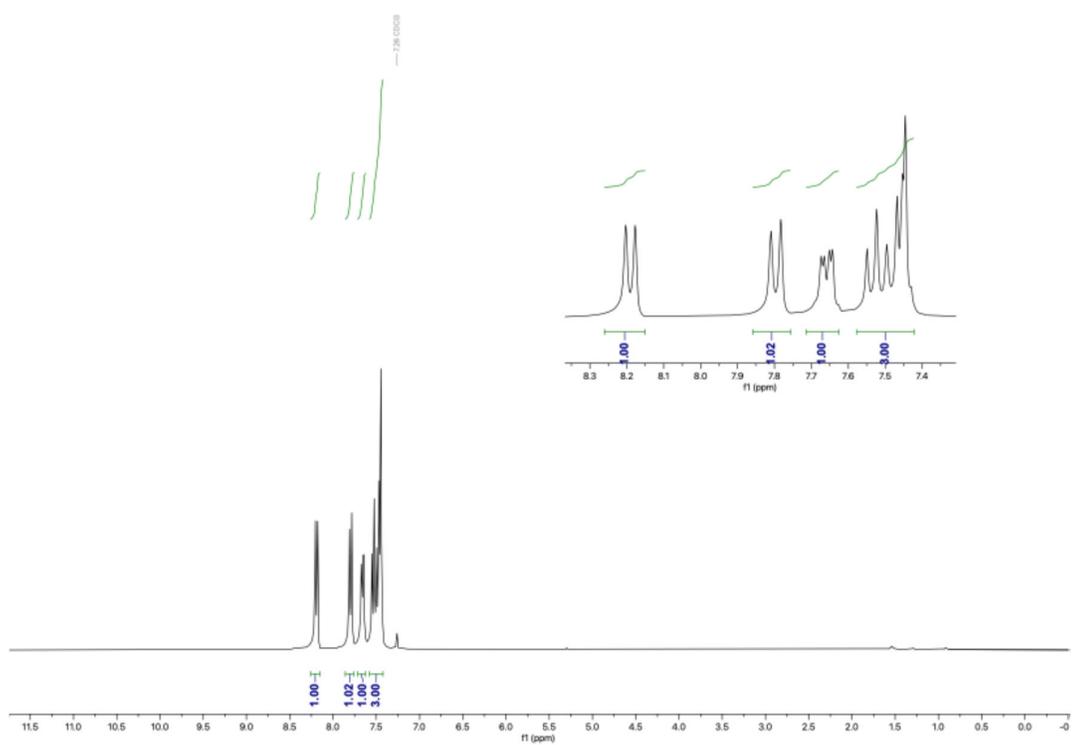

**Figure S7**. $^1$H NMR (300 MHz, CDCl$_3$) spectrum of **DNO**.

## 1.2. Sample preparation and scanning probe microscopy measurements

STM and AFM measurements were performed in a home-built system operating at base pressures below $1 \times 10^{-10}$ mbar and a base temperature of 5 K. Bias voltages were applied to the sample with respect to the tip. All STM and AFM measurements were performed with Cu-coated PtIr tips functionalized with a single carbon monoxide molecule at the tip apex. AFM measurements were performed in non-contact mode with a qPlus sensor.[2] The sensor was operated in frequency-modulation mode[3] with a constant oscillation amplitude of 0.5 Å. STM measurements were performed in constant-current mode, and AFM measurements were performed in constant-height mode with $V = 0$ V. STM and AFM images were post-processed using Gaussian low-pass filters.

The Cu(111) surface was prepared by multiple cycles of sputtering with Ne$^+$ ions and annealing up to 773 K. NaCl was thermally evaporated on the Cu(111) surface held at 283 K. This protocol resulted in the growth of large, defect-free and predominantly bilayer (100)-terminated NaCl films, with a minority of third-layer NaCl islands. The sample quality was ensured by STM imaging before further preparation. Submonolayer coverage of **DNO** on the surface was obtained by flashing an oxidized silicon wafer containing **DNO** molecules in front of the cold sample in the microscope. Carbon monoxide molecules (for tip functionalization) were dosed from the gas phase on the cold sample. The maximum sample temperature during deposition of **DNO** and carbon monoxide molecules was 13 K.

Tip-induced chemistry was performed by relocating the tip at the center of **DNO** molecules at an STM set-point of $V = 0.2$ V and $I = 0.5$ pA. The feedback loop was then opened, and the tip was retracted by 9–11 Å to limit the tunneling current (typically, $I < 50$ pA at the final voltage). The voltage was then ramped from 0.2 V to 4.9–5.1 V in 30 ms and maintained at the final value for 140 ms before ramping back to 0.2 V in 30 ms. The area was scanned after application of the voltage pulses to monitor the occurrence of reactions. Note that molecules were always displaced on the surface during voltage pulses.

## 1.3 DFT calculations

Spin-polarized density functional theory calculations were performed using the Vienna ab initio Simulation Package (VASP).[4–7] The interaction between the core and valence electrons was described using the frozen core projector augmented wave (PAW) method.[8,9] The plane waves were truncated at an energy cutoff of 450 eV in the expansion of the Kohn-Sham orbitals. The considered valence electrons were $1s^1$ (H), $2s^2 2p^2$ (C), $2s^2 2p^4$ (O), $2p^6 3s^1$ (Na), $3s^2 3p^5$ (Cl) and $4s^1 3d^{10}$ (Cu). The exchange-correlation functional was described by the Perdew-Burke-Ernzerhof (PBE) functional.[10] Additional calculations employing the B3LYP functional[11] was used for energy benchmarking.

Grimme's D3 correction was included to account for the dispersion interactions for molecules adsorbed on surfaces.[12,13] The electronic structure was regarded as converged when changes in the electronic energy and Kohn-Sham eigenvalues were below $1 \times 10^{-6}$ eV between successive iterations. The structures were optimized using the conjugate gradient method, until the maximum force was below 0.03 eV/Å.

The lateral lattice constant of a bilayer NaCl was determined to be 5.526 Å using a $(\sqrt{2} \times \sqrt{2})$R45° surface cell (composed of 4 Na and 4 Cl atoms). The lattice constant of bulk Cu was determined to be 3.57 Å, which is slightly lower than the experimental value of 3.61 Å.[14] Gas-phase species were optimized using a (32, 31, 30) Å vacuum box. Transition state calculations were performed using the climbing-image nudged elastic band method[15,16] using a (19, 18, 15) Å vacuum box. Structural optimization and nudged elastic band calculation for anionic species were performed by adding an excess electron to the system. Vibrational analyses were performed assuming the harmonic approximation and solved using finite difference to confirm the transition states in the main text.

The charge state of the system was analyzed using Bader charge analyses.[17] Molecular orbitals of gas-phase molecules were calculated using Dmol$^3$ (ref.[18]).

The investigated surface was composed of a bilayer of $(5\sqrt{2} \times 4\sqrt{2})R45°$ NaCl(100) supported on $(11 \times 5\sqrt{3})$rect. Cu(111). The optimized lattice constant for the bilayer of NaCl was used, whereas the Cu surface was strained to match the lattice of NaCl. The strains on Cu(111) were –0.50 % and 1.11 % in the two dimensions, respectively. The surface is shown in Fig. S8. In the calculations, the bottom layer of the strained Cu(111) was kept fixed to emulate a bulk system.

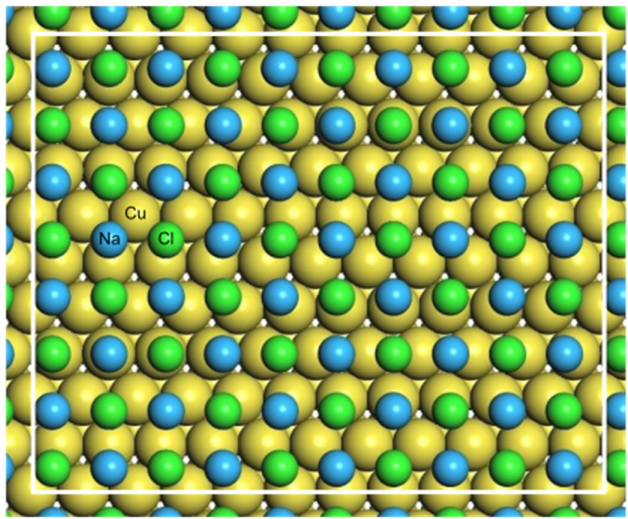

**Figure S8.** Ball model of the investigated bilayer NaCl/Cu(111) system. Atomic color codes: blue (Na), green (Cl), and yellow (Cu).

The main approximation in the DFT calculations is the choice of the exchange-correlation functional. The applied PBE functional is generally a good compromise when having systems with molecules adsorbed on surfaces. However, PBE is known to over-delocalize the electron density, which could influence the energetic preferences. B3LYP is a computationally expensive hybrid functional that reduces the over-delocalization issue. We performed a set of benchmark calculations with the B3LYP functional. The energy difference between **DNO** and **Int3** is –1.16 eV with PBE and –1.11 eV with B3LYP. The energy difference between **Int3** and (Compound **2** + OH) is 4.81 eV with PBE and 4.48 eV with B3LYP. Thus, the trends between the two functionals are consistent.

## 2. Supporting data

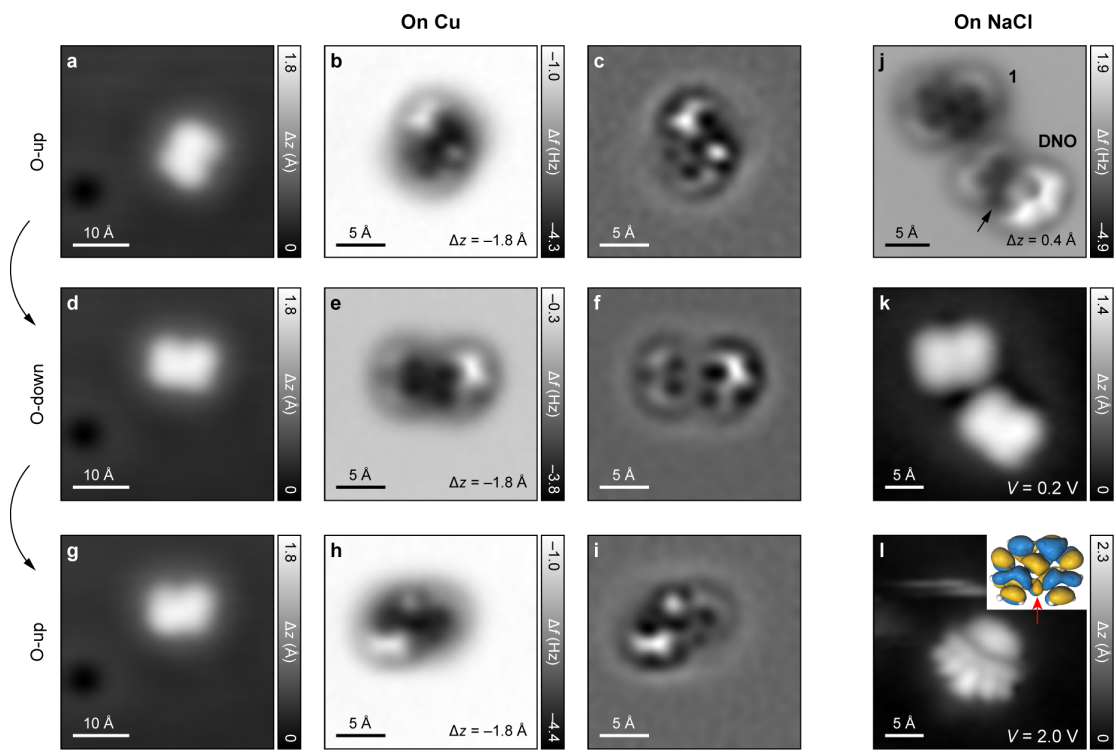

**Fig. S9.** Additional measurements on **DNO**. (a–c) STM (a), AFM (b) and corresponding Laplace-filtered AFM (c) images of an O-up **DNO** molecule on Cu. The oxygen atom appears as a bright protrusion in AFM imaging, as for an O-up **DNO** molecule on NaCl (Fig. 2). (d–f) STM (d) and AFM (e, f) images of an O-down **DNO** molecule on Cu, obtained after application of a voltage pulse of 4.9 V to the O-up molecule in (a–c). The oxygen atom in the O-down conformation is not visible in AFM imaging, as for an O-down **DNO** molecule on NaCl (j and Fig. 2). On further scanning the O-down molecule, its conformation changed back to O-up, as shown in the STM and AFM images in (g–i). (j, k) AFM (j) and STM (k) images of a **1⁻** and an O-down **DNO** molecule adsorbed next to each other (also shown in Fig. 2). (l) Corresponding STM image showing the LUMO density of the **DNO** molecule. The DFT calculated LUMO of **DNO** is also shown (isosurface: $0.01 a_0^{-3/2}$). At this elevated voltage, the **1⁻** molecule was mobile and could not be imaged stably (resulting in the streaks in the image). The arrows in (j, l) indicate the location of the oxygen atom. Scanning parameters for STM images: $V$ = 0.2 V, $I$ = 0.5 pA (a, d, g); $I$ = 0.3 pA (k) and $I$ = 0.2 pA (l). Open feedback parameters for AFM images: $V$ = 0.2 V, $I$ = 0.5 pA on Cu (b, c, e, f, h, i) and NaCl (j).

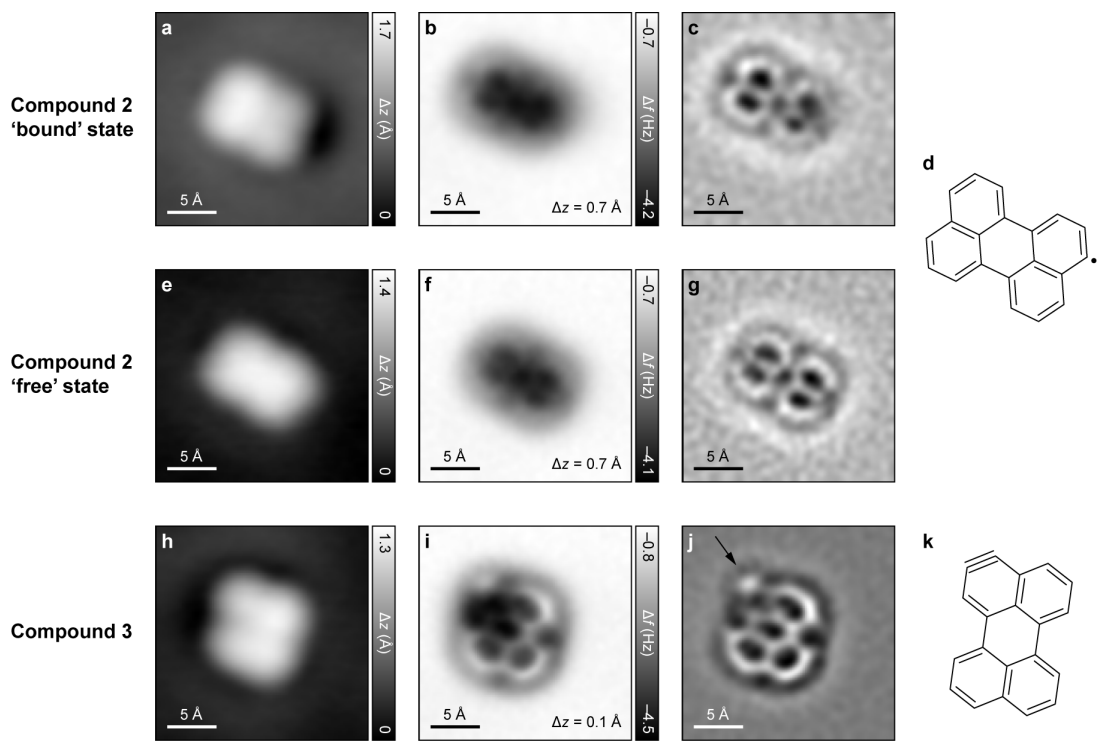

**Fig. S10.** Imaging of compounds **2** and **3**. (a–g) STM and AFM images of a perylenyl radical $C_{20}H_{11}$ (**2**), observed in 2/21 cases. Note that in perylenyl radical, the σ radical may be located at one of the three inequivalent ortho (2, 5, 8, 11 positions), peri (3, 4, 9, 10 positions) or bay (1, 6, 7, 12 positions) sites of perylene. For simplicity, we use the label **2** to denote species with the radical at any of the three sites. We observed **2** to exist in 'free' (highly mobile on the surface) and 'bound' (comparatively less mobile) states on NaCl, as previously observed by Zhong et al. for other polycyclic conjugated hydrocarbon σ radicals.[19] (a, e) STM images of bound (a) and free (e) states of **2**. (b, f) Corresponding AFM images of bound (b) and free (f) states of **2**. (c, g) Laplace-filtered versions of b (c) and f (g). The data in (a–g) were acquired on the same molecule with the same tip. The chemical structure of **2** corresponding to (a–c) and (e–g) is shown in (d). (h–j) STM (h), AFM (i) and corresponding Laplace-filtered AFM (j) images of didehydroperylene $C_{20}H_{10}$ (**3**), observed in 1/21 cases. The bright feature in the upper left benzenoid ring, indicated by an arrow in (j), is related to the C–C triple bond.[20,21] (k) Chemical structure of **3**. Scanning parameters for STM images: $V = 0.2$ V, $I = 0.3$ pA (a, e) and 0.5 pA (h). Open feedback parameters for AFM images: $V = 0.2$ V, $I = 0.5$ pA on NaCl.

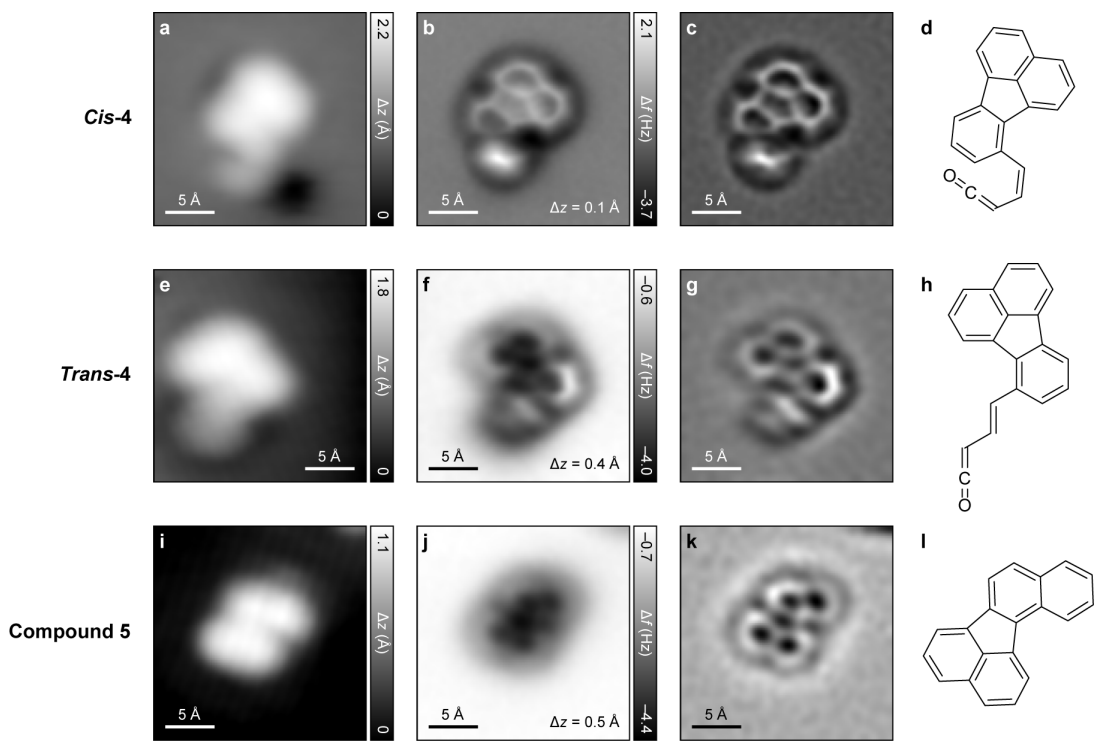

**Fig. S11.** Imaging of compounds **4** and **5**. (a–h) STM and AFM images of *cis* (a–d) and *trans* (e–h) isomers of **4** (observed in 4/21 cases), resulting from ring opening reactions of **DNO**. (a, e) STM images of *cis*- (a) and *trans*- (e) **4**. (b, f) Corresponding AFM images of *cis*- (b) and *trans*- (f) **4**. (c, g) Laplace-filtered versions of b (c) and f (g). (d, h) Tentative chemical structures of *cis*- (d) and *trans*- (h) **4**. (i–k) STM (i), AFM (j) and corresponding Laplace-filtered AFM (k) images of **5** (observed in 1/21 cases). (l) Chemical structure of **5**. Scanning parameters for STM images: $V$ = 0.2 V, $I$ = 0.5 pA. Open feedback parameters for AFM images: $V$ = 0.2 V, $I$ = 0.5 pA on NaCl.

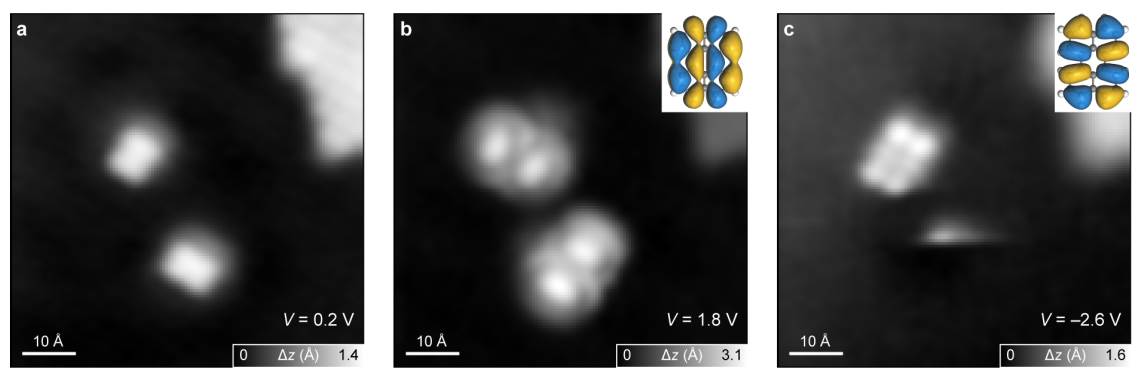

**Fig. S12.** Orbital density imaging of perylene. (a) In-gap STM image of two perylene molecules on NaCl. The upper right corner of the scan frame contains a third-layer NaCl island. (b, c) Corresponding STM images at the voltages indicated in the respective panels, showing the LUMO (b) and HOMO (c) density of perylene. The DFT calculated LUMO and HOMO of perylene are also shown in (b) and (c), respectively (isosurface: $0.01a_0^{-3/2}$). The elevated voltage in (c) led to movement of one of the perylene molecules. Scanning parameters: $I$ = 0.50 pA (a), $I$ = 0.15 pA (b) and $I$ = 0.12 pA (c).

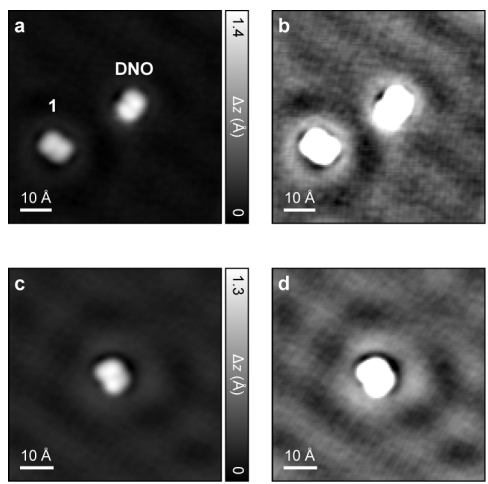

**Fig. S13.** Observation of NaCl/Cu(111) interface-state scattering by compound **1**. (a, b) STM image of **1** and **DNO** shown with two contrast levels. Scattering of the NaCl/Cu(111) interface state[22,23] is observed by **1**, supporting its charged state, while **DNO**, which is neutral, does not scatter the interface state. (c, d) STM image of **1** shown with two contrast levels. Concentric ring-like features are visible around **1** due to interface-state scattering. Scanning parameters: $V$ = 0.2 V (a, b) and $V$ = 0.1 V (c, d), $I$ = 0.5 pA. The voltages in (a–d) are chosen to lie above the onset[22] of NaCl/Cu(111) interface state (at $V \sim -0.2$ V).

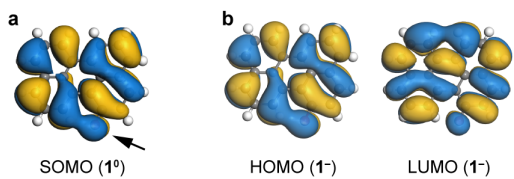

**Fig. S14.** The DFT calculated SOMO of $\mathbf{1^0}$ (a), and HOMO (left) and LUMO (right) of $\mathbf{1^-}$ (b) (isosurface: $0.01a_0^{-3/2}$). $\mathbf{1^0}$ is an open-shell molecule with an unpaired π-electron, whereas $\mathbf{1^-}$ is a closed-shell molecule. The arrow in (a) indicates the oxygen atom.

**Table S1.** Optimized C−O bond lengths of selected molecules from DFT calculations.

| Molecule | Charge state | C−O bond length (Å) |
|---|---|---|
| Compound **1** 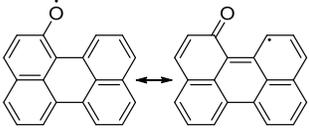 | Neutral, in gas phase<br>Anionic, in gas phase<br>Anionic, on bilayer NaCl/Cu(111) | 1.26<br>1.27<br>1.29 |
| **Int3** 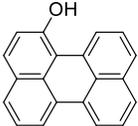 | Neutral, in gas phase<br>Anionic, in gas phase<br>Neutral, on bilayer NaCl/Cu(111) | 1.38<br>1.40<br>1.37 |
| Cyclohexanone 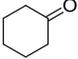 | Neutral, in gas phase<br>Anionic, in gas phase | 1.23<br>1.25 |
| Phenol 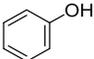 | Neutral, in gas phase<br>Anionic, in gas phase | 1.38<br>1.37 |

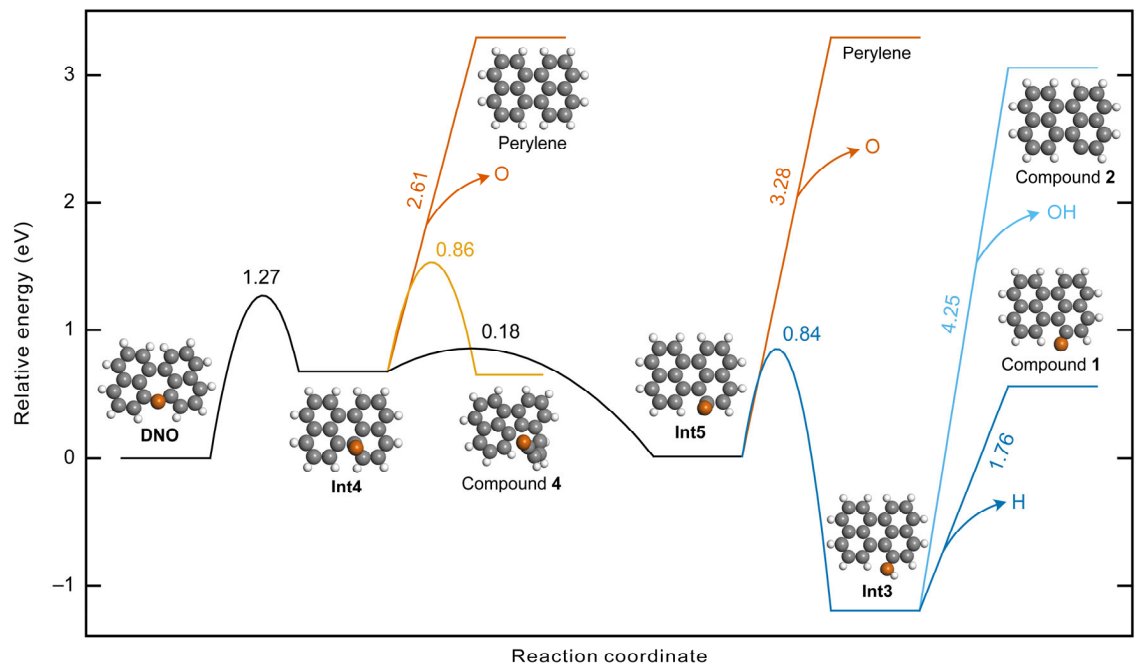

**Fig. S15.** Calculated gas-phase potential energy landscape for skeletal editing of **DNO**, assuming a global anionic state; that is, anionic states of the reactant, intermediates and products. The optimized gas-phase geometries of the molecular species are also shown. The numbers adjacent to the curves denote activation energies in eV. Note that for the reactions **Int4** and **Int5** → perylene and **Int3** → compounds **1** and **2**, energy differences coincide with activation energies. All activation energies are lowered compared to the neutral case (Fig. 3 in the main text). In the intermediates **Int4** and **Int5**, the oxygen atom is located on top of a carbon atom. For the neutral case, the corresponding intermediates **Int1** and **Int2** have the oxygen atom located in a C–C bridge site (Fig. 3).

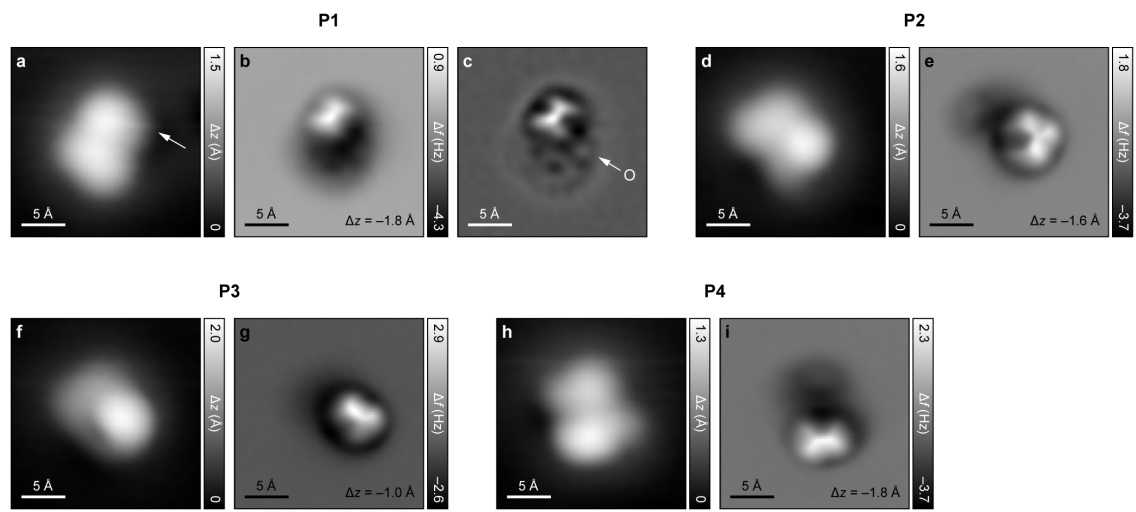

**Fig. S16.** Reaction products on Cu(111). STM and AFM images of four products **P1–P4** obtained after application of voltage pulses of $V = 5.0$ V (**P1–P3**) and $V = 4.9$ V (**P2**) to individual **DNO** molecules. (a–c) STM (a), AFM (b) and Laplace-filtered AFM (c) images of **P1**, which still contains the oxygen atom indicated by the arrow in (c). The benzenoid ring, whose location is indicated by the arrow in (a), has likely lost one or more hydrogen atoms via cleavage of C($sp^2$)–H bonds. This ring is not resolved in AFM imaging (b, c) likely due to the formation of C–Cu bonds, which leads to bending of the ring toward the surface. (d–i) STM (d, f, h) and AFM (e, g, i) images of **P2**–**P4**, which are difficult to identify. It is likely that **P2**–**P4** exhibit loss of one or more hydrogen atoms. However, it is unclear if the oxygen atom is present in these molecules. Scanning parameters for STM images: $V$ = 0.2 V, $I$ = 0.5 pA. Open feedback parameters for AFM images: $V$ = 0.2 V, $I$ = 0.5 pA on Cu.